# Sensitivity of Collective Plasmon Modes of Gold Nanoresonators to Local Environment

V. G. Kravets[1], F. Schedin[1], A. V. Kabashin[2] and A. N. Grigorenko[1]

[1]*School of Physics and Astronomy, University of Manchester, Manchester, M13 9PL, UK.*

[2]*Laboratoire Lasers, Plasmas et Procédés Photoniques (LP3 UMR 6182 CNRS), Faculté des Sciences de Luminy, Université de Méditerranée, Case 917, 13288 Marseille, France.*

**Abstract.**

We present the first experimental study of optical response of collective plasmon resonances in regular arrays of nanoresonators to local environment. Recently observed collective plasmon modes arise due to diffractive coupling of localised plasmons and yield almost an order of magnitude improvement in resonance quality. We measure the response of these modes to tiny variations of the refractive index of both gaseous and liquid media. We show that the phase sensitivity of the collective resonances can be more than two orders of magnitude better than the best amplitude sensitivity of the same nanodot array as well as an order of magnitude better than the phase sensitivity in SPR sensors.

A high sensitivity of optical excitations of the electron plasma in metal films and nanostructures (referred to as surface plasmons [1, 2]) to the refractive index (RI) of local environment is widely applied for development of sensitive bio- and chemical sensors. For example, surface plasmon resonance (SPR) observed in the Kretschmann-Raether configuration has been made to become a benchmark in biosensing by Biacore. The sensitivity of SPR technique in the visible range is of the order of 1 nm per $10^{-3}$ of refractive index unit (RIU), which makes possible the achievement of the detection limit of $10^{-5}$ – $10^{-6}$ RIU [3]. This sensitivity can be enhanced by using phase of light under SPR conditions [4] which was shown to demonstrate dramatic changes near the SPR minimum of reflection and allowed one to measure local RI changes at the level of $10^{-8}$ RUI [5, 6, 7]. However, the excitation of SPR over a thin film does not always satisfy modern requirements of nano-biotechnology [8] which advances toward novel nanoscale architectures, including beacons [9] and other nano-molecular complexes, and demands new functionalities such as manipulation on a nanoscale level [10], size-based selectivity and multi-sensing in microarrays with single particle sensitivity [11]. Localised plasmon resonances (LPR) observed in metallic nanostructures [12] represent a viable alternative to SPR based on propagating plasmons for the tasks of bio- and chemical sensing. LPR appear to be much more suitable to match these new trends, as well as to bring new properties, such as spectral tuneability, strong enhancement of the local electric field, and nano-trapping. For example, LPR of gold double-pillar nanomolecules working in visible light can be tuned in a 100nm wavelength range just by changing polarization of incident light [13, 14]. However, LPR-based sensors are known to provide orders of magnitude lower sensing response to refractive index change compared to SPR with sensitivities not exceeding 0.1-0.3 nm per $10^{-3}$ RIU in spectral interrogation schemes [12, 15, 16, 17].

Recently, it was suggested theoretically [18, 19] and shown experimentally [20, 21] that regular arrays of metallic nanostructures possess another type of extremely narrow collective plasmon resonances (CPR) which are based on diffractive coupling of localised plasmons. These resonances can be explained as Wood anomalies [20] and are quite sensitive to local environment. For example, the sensitivity of the "air" CPR [20] observed at 600nm can be evaluated as 0.6nm per $10^{-3}$ RIU which is comparable with the experimental sensitivity of SPR. In addition, it was shown [20] that a suitable choice of sizes of gold nanoinclusions allows one to achieve a compete suppression of light reflection at the resonance conditions that can be useful for the phase methods of detection [5, 6, 7]. In this Letter we assess an application of CPR for bio- and chemical sensing based on RI monitoring and show that CPR holds a big promise for developments of a new generation of plasmonic sensors.

To observe CPR we have used regular arrays of Au nanodots (nanopillars), as shown in the Inset of Fig. 1, fabricated by high-resolution electron-beam lithography on a glass substrate [10, 14, 20]. Optical characteristics from the nanodot arrays were examined in direct and attenuated total reflection geometries (see top insets in Fig. 1), corresponding to gas and bio- sensing, respectively, with a focused beam Woollam ellipsometer. The resonance is plotted in $\Psi$-$\Delta$ values routinely used in ellipsometry so that $E_p / E_s = \tan(\Psi)\exp(i\Delta)$ , where $E_p$ and $E_s$ are the reflected field amplitudes for the incident light $E_i$ of *p*- and *s*-polarizations respectively. A deep and narrow CPR for a single dot array was observed at $\theta_{inc}$=71° in the red region of spectra for the square array of single Au nanoparticles with dot diameter *d*=118nm and lattice constant *a*=316nm, see Fig. 1(a). The observed CPR has a narrow half-width of just $\Delta\lambda_{HW}$=6nm and yields almost totally suppressed reflection of *p*-polarization $R_p$≈0 at $\lambda_{min}$=613nm. The resonance wavelength in the studied sub-wavelength structures is close to the "air" Rayleigh cut-off wavelength [20] $\lambda_{R_1}^{air} = n_g a(1+\sin(\theta_{inc}))$, where $n_g$ is the RI of the gas surrounding nanodots ($\lambda_{R_1}^{air}$≈615nm for the single dot array of Fig. 1(a)). Figure 1(c) shows a deep and narrow CRP observed at $\theta_{inc}$=63° for the square array of double-pillar Au "nanomolecules" with *d*=100nm, *a*=317nm and a pillar separation of *s*=140nm. The observed CPR has a narrow halfwidth $\Delta\lambda_{HW}$=5nm and also demonstrates almost totally suppressed reflection of *p*- polarized light ($R_p \approx 0$) at the resonance wavelength $\lambda_{min}$=598nm. For ATR geometry and water solutions used in bio-sensing, CPR of our arrays shifted toward longer wavelengths. It is easy to show that the strongest resonance for ATR configuration with a $\pi$/4-prism ($n_{pr}$=1.5) can be found as $\lambda_R^{ATR} = a\left(n_{pr}\sin\left(\pi/4\right) + \arcsin\left(\sin(\theta_{inc} - \pi/4)/n_{pr}\right) + \sin(\theta_{inc} - \pi/4)/\sqrt{2} + n_{liq}\right)$, where $n_{liq}$ is the refractive index of studied liquid. For $a$=320nm, $\theta_{inc}$=45° and $n_{liq}$=1.33 this gives $\lambda_R^{ATR}$ =765nm. To realise deep CPR one needs to match LPR of "nanomolecules" with the Rayleigh cut-off wavelength [20]. This was achieved at the following array parameters $d$=132nm, $a$=318nm, s=140nm in ATR configuration at $\theta_{inc}$=45°, see Figs. 1(e). (Relevant LPR of double pillars are described in [14].) Figure 1(e) show a deep and narrow resonance at $\lambda_{min} \approx$764nm with a half-width of just $\Delta\lambda_{HW}$=5nm. The double pillar structure in ATR geometry also demonstrates the condition of almost totally suppressed reflection of light of $p$-polarization at $\lambda_{min}$=764nm. It is worth noting extremely fast changes of light phase $\Delta$ near the condition of $R_p \approx 0$, see Fig. 1(b), (d) and (f). These changes are analogous to the behaviour of light reflection from a dielectric material near the Brewster angle or gold periodic gratings [22, 23] and important for realization of high phase sensitivity [6, 20, 24].

To study the response of our structures to refractive index of environment we have used controlled gaseous mixtures of propanol with air and controlled mixtures of water with glycerol. Figure 2 and Table 1 show the main result of this Letter - the amplitude and phase responses of the described samples, as well as theoretical data for SPR in the Kretschmann-Raether geometry for comparison. In the case of the direct gas measurements, we applied two different gas mixtures and plotted the difference of the amplitude signal, $\delta\Psi$, and the phase signal, $\delta\Delta$, in the region near CPR. Figure 2(a) plots the difference signal for the double dot array of Figs. 1(c), (d) that corresponds to two different propanol-air mixtures with $\Delta n = 4\times10^{-5}$. The results for the ATR geometry for array of Figs. 1(e), (f) are shown in Fig. 2(b) for the distilled water and mixture of distilled water with glycerol with the difference in refractive index of $\Delta n=6\times10^{-4}$. It is worth noting that the data of Fig. 2 (and Fig. 1) represent raw data without any averaging.

From Table 1 we see that CPR sensitivity to local index of environment in the gaseous media has an edge over the SPR in the Kretschmann-Raether configuration. Indeed, despite the fact the CPR wavelength shift is smaller in the studied samples than in a theoretical case of SPR, the corresponding amplitude signals are comparable with those of theoretical SPR (at the level of 3° of $\Psi$ per $10^{-3}$ RIU which corresponds to about 10% in light intensity). This is connected to the better resonance quality of CPR in comparison to SPR [20]. The resonance quality can be roughly evaluated as $Q=\lambda_{min}/\Delta\lambda_{HW}$. The experimental CPR observed in the studied samples has quality of $Q\approx100$ in the air and $Q\approx150$ for ATR geometry which is an order of magnitude better than that of conventional SPR ($Q$~10). However, the biggest improvement of CPR sensing scheme lies in its extremely high phase sensitivity. From Figure 2 and Table 1 we see that the best experimental CPR phase sensitivity in gaseous media (520° per $10^{-3}$ RIU) is more than 2 orders of magnitude better than the corresponding amplitude sensitivity and one order of magnitude better than the phase sensitivity of SPR [5]. Simple considerations show that the phase sensitivity of a resonance curve depends on the value of the intensity of reflected light at the resonance minimum [6]. Introducing a figure of merit, FOM, as a ratio of the phase over the amplitude sensitivity [24], we get $\mathrm{FOM} = \delta\Delta / \delta\Psi \sim 1/\Psi_{\min}$, where $\Psi_{min}$ is expressed in radians. (Neglecting the change of reflection for *s*-polarization, we have the phase changes near the resonance minima as $\delta\Delta \sim \delta E_{p,\min} / E_{p,\min}$ and the amplitude changes at the slope of the resonance as $\delta\Psi \sim \delta E_{p,sl} / E_{p,sl}$. Taking into account that $\delta E_{p,\min} \sim \delta E_{p,sl}$ and $E_{p,sl} \sim E_{s,\min}$, we obtain the expression above). Thus, low values of the minimum of reflection at CPR ($\Psi_{min}\approx0.3°$) guarantee its extremely high phase sensitivity (FOM~200). Assuming an accuracy of phase measurements in a scheme with fixed angle of incidence at a level of about 0.001° [3], we can evaluate the detection limit for sensors based on phase CPR as $2\times10^{-9}$ RIU, which overcomes that one for phase-sensitive SPR counterparts by almost an order of magnitude [24]. By changing array parameters (e.g., the lattice constant *a*) it is realistic to develop a phase sensitive CPR sensor with the threshold at the level of $10^{-10}$ RIU.

To conclude, we have shown that CPR in regular arrays of gold nanoresonators can be a viable alternative to SPR for development of chemical and bio-sensors. We demonstrated that experimental amplitude sensitivity of CPR to local index of refraction is comparable with that of SPR, while CPR phase sensitivity is an order of magnitude better than that of SPR based on the Kretschmann configuration. The interrogation spot of our experiments was about 60x30 $\mu m^2$ which allow one to develop high-through-put and chip-based detection techniques.

**Acknowledgements**

This work has been supported by Paul Instrument Grant and EPSRC grant EP/E01111X/1. We are grateful to the Reviewer who indicated a link between our results and the total absorption of light observed on periodic gratings [22, 23].

**References.**

1. R. H. Ritchie, "Plasma Losses by Fast Electrons in Thin Films," *Phys. Rev.* **106**, 874-881 (1957).

2. H. Raether, Surface plasmons, in *Springer Tracts in Modern Physics* (Springer-Verlag, 1988), Vol. **111**.

3. J. Homola, S. S. Yee, and G. Gauglitz, "Surface plasmon resonance sensors: review," *Sensors and Actuators B: Chemical* **54**, 3-15 (1999).

4. F. Abeles, "Surface electromagnetic waves ellipsometry," *Surface Science* **56**, 237-251 (1976).

5. A. V. Kabashin and P. I. Nikitin, "Interferometer based on a surface-plasmon resonance for sensor applications," *Quantum Electronics* **27**, 653-654 (1997).

6. A. N. Grigorenko, P. I. Nikitin, and A. V. Kabashin, "Phase jumps and interferometric surface plasmon resonance imaging," *Applied Physics Letters* **75**, 3917-3919 (1999).

7. P. I. Nikitin, A. N. Grigorenko, A. A. Beloglazov, M. V. Valeiko, A. I. Savchuk, O. A. Savchuk, G. Steiner, C. Kuhne, A. Huebner, and R.; Salzer, "Surface plasmon resonance interferometry for micro-array biosensing," *Sensors and Actuators A-Physical* **85**, 189-193 (2000).

8. P. N. Prasad, *Introduction to Biophotonics* (Wiley-Interscience, New Jersey, 2003).

9. M. B. Wabuyele and T. Vo-Dinh, "Detection of Human Immunodeficiency Virus Type 1 DNA Sequence Using Plasmonics Nanoprobes," *Analytical Chemistry* **77**, 7810-7815 (2005).

10. A. N. Grigorenko, N. W. Roberts, M. R. Dickinson, and Y. Zhang, "Nanometric optical tweezers based on nanostructured substrates," *Nat Photon* **2**, 365-370 (2008).

11. M. Schena, D. Shalon, R. W. Davis, and P. O. Brown, "Quantitative Monitoring of Gene Expression Patterns with a Complementary DNA Microarray," *Science* **270**, 467-470 (1995).

12. M. D. Malinsky, K. L. Kelly, G. C. Schatz, and R. P. Van Duyne, "Chain Length Dependence and Sensing Capabilities of the Localized Surface Plasmon Resonance of Silver Nanoparticles Chemically Modified with Alkanethiol Self-Assembled Monolayers," *Journal of the American Chemical Society* **123**, 1471-1482 (2001).

13. W. Rechberger, A Hohenau, A. Leitner, J. R. Krenn, B. Lamprecht, F. R. Aussenegg, "Optical properties of two interacting gold nanoparticles," *Optics Communications* **220**, 137-141 (2003).

14. A. N. Grigorenko, A. K. Geim, H. F. Gleeson, Y. Zhang, A. A. Firsov, I. Y. Khrushchev, J. Petrovic, "Nanofabricated media with negative permeability at visible frequencies," *Nature* **438**, 335-338 (2005).

15. K. Su, Q. Wei, X. Zhang, J. J. Mock, D. R. Smith, S. Schultz, "Interparticle Coupling Effects on Plasmon Resonances of Nanogold Particles," *Nano Letters* **3**, 1087-1090 (2003).

16. A. N. Grigorenko, H. F. Gleeson, Y. Zhang, N. W. Roberts, A. R. Sidorov, A. A. Panteleev, "Antisymmetric plasmon resonance in coupled gold nanoparticles as a sensitive tool for detection of local index of refraction," *Applied Physics Letters* **88**, 124103 (2006).

17. J. N. Anker, W. P. Hall, O. Lyandres, N. C. Shah, J. Zhao, R. P. Van Duyne, "Biosensing with plasmonic nanosensors," *Nat Mater* **7**, 442-453 (2008).

18. S. Zou, N. Janel, and G. C. Schatz, "Silver nanoparticle array structures that produce remarkably narrow plasmon lineshapes," *The Journal of Chemical Physics* **120**, 10871-10875 (2004).

19. V. A. Markel, "Divergence of dipole sums and the nature of non-Lorentzian exponentially narrow resonances in one-dimensional periodic arrays of nanospheres," *Journal of Physics B: Atomic, Molecular and Optical Physics* **38**, L115-L121 (2005).

20. V. G. Kravets, F. Schedin, and A. N. Grigorenko, "Extremely Narrow Plasmon Resonances Based on Diffraction Coupling of Localized Plasmons in Arrays of Metallic Nanoparticles," *Physical Review Letters* **101**, 087403 (2008).

21. B. Auguie and W. L. Barnes, "Collective Resonances in Gold Nanoparticle Arrays," *Physical Review Letters* **101**, 143902 (2008).

22. M. C. Hutley and D. Maystre, "Total absorption of light by a diffraction grating," *Optics Communications* **19**, 431-436 (1976).

23. E. Popov, D. Maystre, R. C. McPhedran, M. Neviere, M. C. Hutley, G. H. Derrick, "Total absorption of unpolarized light by crossed gratings," *Optics Express* **16**, 6146-6155 (2008).

24. A. V. Kabashin, S. Patskovsky, A. N. Grigorenko, "Phase and amplitude sensitivities in surface plasmon resonance bio and chemical sensing", *Optics Express* **17**, 21191-21204 (2009).

**Figure Captions.**

Fig. 1. (Color online)

Top insets: Direct and ATR geometries of experiments and a SEM micrograph of a typical double-dot array. Bottom: Narrow resonances based on diffractive coupling of localised plasmons (CPR). (a) and (b) Ellipsometric parameters $\Psi$ and $\Delta$ for light reflection from a square array of single Au dots. (c) and (d) $\Psi$ and $\Delta$ for a square array of Au double dots. (e) and (f) $\Psi$ and $\Delta$ for a square array of Au double dots (dumbbells), ATR geometry. Insets on the right show SEM micrographs of an elementary cell of an array.

Fig. 2. (Color online)

A typical CPR response to a change in local index of refraction. (a) The difference amplitude signal $\delta\Psi$ and phase signal $\delta\Delta$ for CPR of Fig. 1(c), (d) registered for a change in gas refractive index $\Delta n=4\times10^{-5}$. (b) $\delta\Psi$ and $\delta\Delta$ for CPR of Fig. 1(e), (f) registered for a change in liquid refractive index $\Delta n=6\times10^{-4}$.

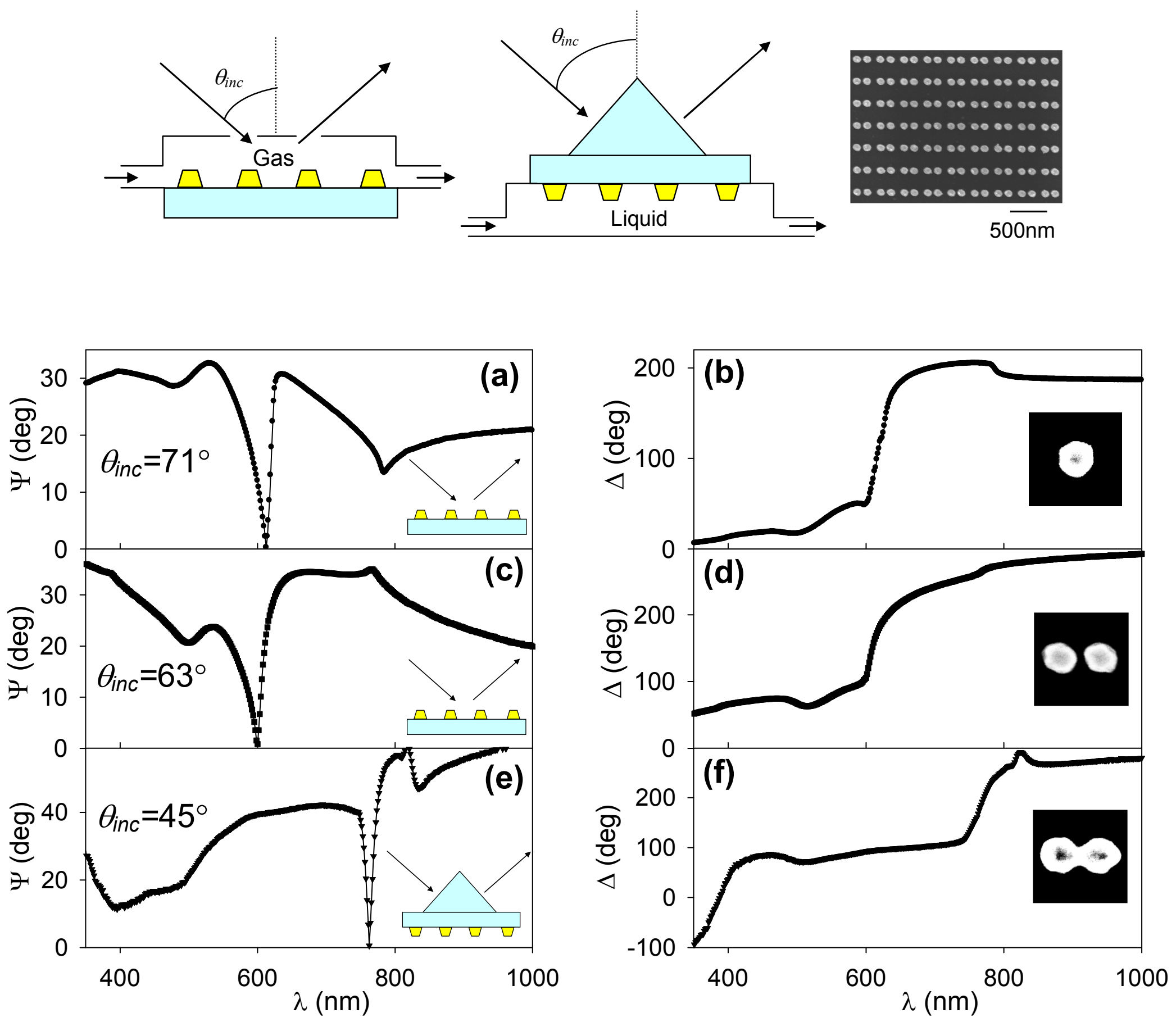

Fig. 1.

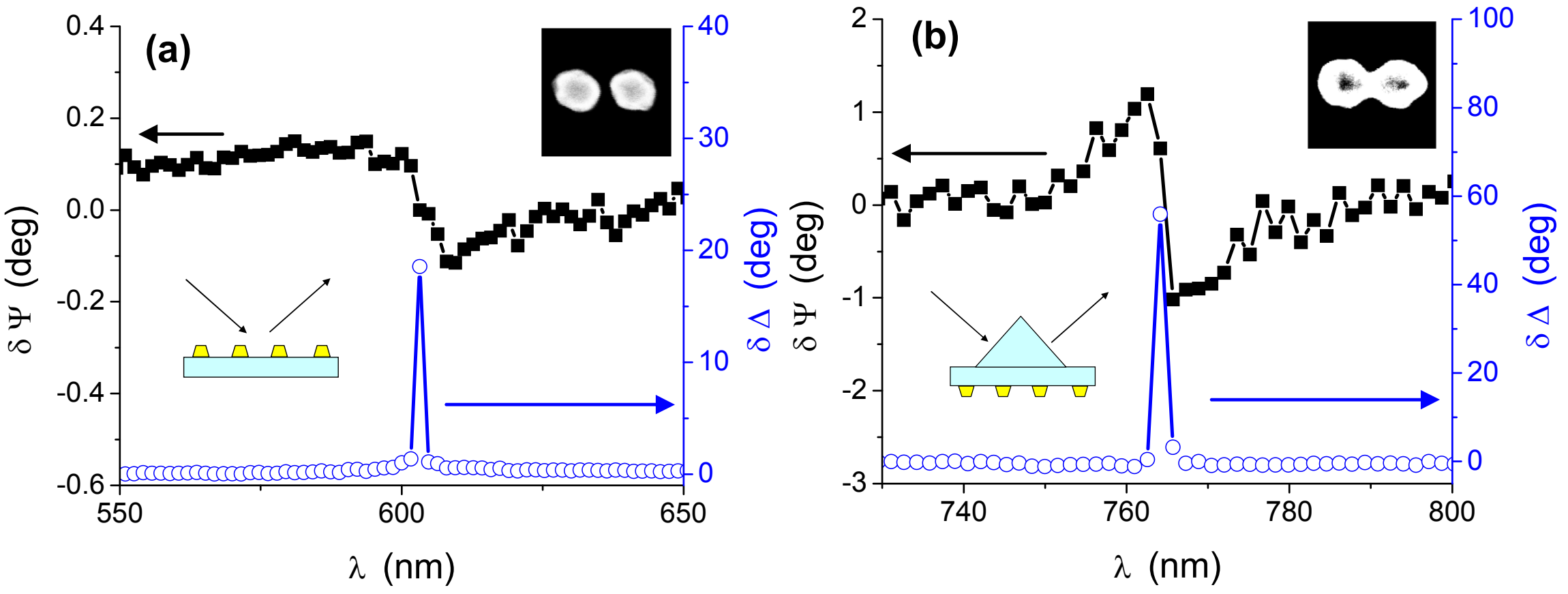

Fig. 2.

Table 1

Amplitude and phase sensitivities of CPR and SPR.

| Sample | Geometry | Sizes | Resonance Position | Amplitude signal at the minimum | Resonance shift per $10^{-3}$ RIU | Amplitude signal per $10^{-3}$ RIU | Phase signal per $10^{-3}$ RIU | FOM, Ratio $\delta\Delta/\delta\Psi$ |
|---|---|---|---|---|---|---|---|---|
| V22TM | Single Dots; Gas ($n\approx1$) | $d$=118nm, $a$=316nm | $\lambda$=613nm $\theta_{inc}$=71° | $\Psi_{min}$=0.29° | 0.62nm | 2.9° | 520° | 180 |
| V3BL8 | Double Dots; Gas ($n\approx1$) | $d$=100nm, $s$=140nm, $a$=317nm | $\lambda$=598nm $\theta_{inc}$=63° | $\Psi_{min}$=0.81° | 0.6nm | 3.2° | 465° | 140 |
| V22TL | Dumbbells; Liquid ($n\approx1.33$) | $d$=132nm, $s$=140nm $a$=318nm | $\lambda$=764nm $\theta_{inc}$=45° | $\Psi_{min}$=0.92° | 0.3nm | 1.85° | 93° | 50 |
| Theory | SPR; n=1.5 prism Gas ($n\approx1$) | 43nm Au | $\lambda$=616nm $\theta_{inc}$=46° | $\Psi_{min}$=3.9° | 3nm | 3.1° | 46° | 15 |
| Theory | SPR; n=1.7 prism Liquid ($n\approx1.33$) | 45nm Au | $\lambda$=622nm $\theta_{inc}$=60° | $\Psi_{min}$=6.2° | 1.8nm | 2.6° | 18° | 7 |